\documentstyle{article}
\begin{document}
\def\wick#1#2{\hspace{-#1mm}\raisebox{-2ex}{\rule{0.02mm}{2mm}\rule{#2mm}{0.02mm}\rule{0.02mm}{2mm}}\hspace{#1mm}\hspace{-#2mm}}
% The wick macro requires 2 parameters: The first is how far back the first
% end of the wick contraction should be placed and the second how long it
% should be.
\newcommand{\one}{\hbox{ 1\kern-.8mm l}}
\newcommand{\ie}{{\it i.e.}\ }
\newcommand{\beq}{\begin{equation}}
\newcommand{\eeq}{\end{equation}}
\newcommand{\nn}{\nonumber}
\newcommand{\tr}{{\mathrm{tr}}}
\newcommand{\str}{{\mathrm{str}}}
\def\ii{\'{\char'20}}
\def\r{\rightarrow}
\def\err{\end{array}}
\def\bea{\begin{eqnarray}}
\def\eea{\end{eqnarray}}
\def\bp{{\bf p}}
\def\bk{{\bf k}}
\def\bq{{\bf q}}
\def\ttau{\tilde{\tau}}
\def\tchi{\tilde{\chi}}
\def\trho{\tilde{\rho}}
\def\teps{\tilde{\epsilon}}
\def\tnu{\tilde{\nu}}
\def\tgamma{\tilde{\gamma}}

%%%%%%%%%%%%%%%%%%%%%%%%%%%%% Title
\begin{center}
{\Large \bf Gauge invariant regularisation in the ERG approach} \\

\vspace{4mm}

%%%%%%%%%%%%%%%%%%%%%%%%%%%%% author/address
S. Arnone$^{\;a}$, Yu.A. Kubyshin$^{\,a,b}$, 
T.R. Morris$^{\;a}$ and J.F. Tighe$^{\;a}$ \\
$^{a}$ Department of Physics and Astronomy, University of Southampton, 
Highfield,\\ Southampton SO17 1BJ, UK \\
$^{b}$ Institute for Nuclear Physics, Moscow State University, 
119899 Moscow, Russia\\
\end{center}

\vspace{-5cm}
\begin{flushright}
SHEP 01-04 
\end{flushright}
\vspace{5cm}
%%%%%%%%%%%%%%%%%%%%%%%%%%%% Abstract
\begin{abstract}
A gauge invariant regularisation which can be used for non-perturbative 
treatment of Yang-Mills theories within the 
exact renormalization group approach is constructed. It consists of a 
spontaneously broken $SU(N|N)$ super-gauge extension of the initial 
Yang-Mills action supplied with covariant higher derivatives. We 
demonstrate that the extended theory in four dimensions 
is ultra-violet finite perturbatively 
and argue that it has a sensible limit when the regularisation cutoff 
is removed.  
\end{abstract}

%%%%%%%%%%%%%%%%%%%%%%%%%%% Body of the article       

\section{Introduction}

The exact renormalization group (ERG), which is the broadly accepted name 
for the continuous version of Wilson's renormalization group \cite{WK},
is a quite powerful tool for studies in quantum field theory 
(see, for example, refs \cite{TMo,TM-R1,TM-R2,Ber}). 
It possesses some important features which motivate the work devoted to 
both further developments of the ERG formalism and to practical 
computations by this method. One of the advantageous features it offers is 
that it allows
non-perturbative (though approximate in practice) studies and 
calculations. Another is that within the ERG approach almost all 
approximations preserve a crucial property of quantum field theory, 
namely the existence of the continuum limit. It also gives rise to the
possibility of   
studying (at least within some approximation) 
the full parameter space of non-perturbative quantum 
field theories, its fixed points, continuum limits, etc. The challenge  
is to understand the low energy limit of QCD within the ERG approach 
(for some preliminary work on this see, for example, \cite{Wet1}).   

The main ingredient of the ERG approach is an ERG equation. It determines the 
effective action (running action), $S_{\Lambda}$, as a function of the 
scale $\Lambda$ which also plays the r\^{o}le of the effective 
momentum cutoff. Let us consider a scalar theory in $D$ dimensions 
for simplicity. The effective action can be written as 
the sum of an (effective) action of interaction, $S_{\Lambda}^{int}[\phi]$, 
and the kinetic term, $\hat{S}_{\Lambda}[\phi]$: 
\[
S_{\Lambda}[\phi] = \hat{S}_{\Lambda}[\phi] + S_{\Lambda}^{int}[\phi].
\]
To regularise the theory we modify the propagator in the momentum space, 
$1/p^{2}$, to $c(p^{2}/\Lambda^{2})/p^{2}$, where 
$c (p^{2}/\Lambda^{2})$ is a (smooth) ultraviolet cutoff profile 
satisfying $c(0) = 1$ so that low energies are unaltered, and 
$c(z) \rightarrow 0$ as $z \rightarrow \infty$ sufficiently fast so that 
all Feynman diagrams are ultraviolet regulated.\\  
Introducing the shorthand 
\beq
f \cdot W \cdot g := \int d^{D}x \, f(x) 
W \left( -\frac{\partial^{2}}{\Lambda^{2}} \right) g(x),   \label{fdWdg-def}
\eeq 
we write the regularised kinetic term as  
\beq
\hat{S}_{\Lambda}[\phi] = \frac{1}{2} \partial_{\mu} \phi \cdot 
c^{-1} \cdot \partial^{\mu} \phi = 
\frac{1}{2} \int \frac{d^{D}p}{(2\pi)^{D}} \, \phi (-p) p^{2} 
c^{-1} \left(\frac{p^{2}}{\Lambda^{2}} \right) \phi (p). \label{S-kin}
\eeq

Polchinski's version of the ERG equation \cite{Polch} is 
\beq
\Lambda \frac{\partial S^{int}_{\Lambda}}{\partial \Lambda} = 
- \frac{1}{\Lambda^{2}} 
\frac{\delta S^{int}_{\Lambda}}{\delta \phi} \cdot c' \cdot  
\frac{\delta S^{int}_{\Lambda}}{\delta \phi} + 
\frac{1}{\Lambda^{2}} \frac{\delta}{\delta \phi} \cdot c' \cdot  
\frac{\delta S^{int}_{\Lambda}}{\delta \phi}. 
\eeq
Being supplied with the initial condition 
\[
S_{\Lambda}^{int}|_{\Lambda = \Lambda_{0}} = \tilde{S}^{int},
\]
it determines the renormalization group flow of the effective action 
which corresponds to (effectively) integrating out higher momentum modes 
\cite{WH}. 
For the scalar case the initial value for the action is usually taken 
to be $\tilde{S}^{int} = \frac{1}{4!} \lambda \int d^{D}x \, \phi^{4}(x)$. 
 
To see the physical meaning of Polchinski's equation it is convenient 
to re-write it in terms of the total effective action $S_{\Lambda}[\phi]$ 
and $\Sigma_{1} = S_{\Lambda} - 2\hat{S}_{\Lambda}$. One gets 
\beq
\Lambda \frac{\partial S_{\Lambda}}{\partial \Lambda} = 
- \frac{1}{\Lambda^{2}} 
\frac{\delta S_{\Lambda}}{\delta \phi} \cdot c' \cdot  
\frac{\delta \Sigma_{1}}{\delta \phi} + 
\frac{1}{\Lambda^{2}} \frac{\delta}{\delta \phi} \cdot c' \cdot  
\frac{\delta \Sigma_{1}}{\delta \phi}.  \label{P-eq}
\eeq 
It can be checked that the kinetic term (\ref{S-kin}) is the Gaussian fixed 
point of this equation. 
Eq.\ (\ref{P-eq}) can also be put into the form 
\beq
\Lambda \frac{\partial}{\partial \Lambda} e^{-S_{\Lambda}} = 
-\frac{1}{\Lambda^{2}} \frac{\delta}{\delta \phi} \cdot c' \cdot \left( 
\frac{\delta \Sigma_{1}}{\delta \phi} e^{-S_{\Lambda}} \right), \label{P-eq1} 
\eeq
which shows that the partition function 
${\cal Z}= \int {\cal D}\phi e^{-S_{\Lambda}}$ remains unchanged along 
the renormalization group flow. Indeed, from (\ref{P-eq1}) it 
follows that 
\[
\Lambda \frac{\partial}{\partial \Lambda} {\cal Z} \equiv 
\Lambda \frac{\partial}{\partial \Lambda} \int {\cal D}\phi e^{-S_{\Lambda}} =
0.  
\]

A number of versions of the ERG equation for the scalar theory have been known 
for more than 20 years \cite{WH,Polch,HH,TM0,Wet2}. 
A Polchinski-type ERG equation for fermions was derived and 
studied in ref.\ \cite{CKM}. In previous ERG approaches to gauge 
theory the authors kept the gauge fixed and allowed the effective 
momentum cutoff to break the gauge invariance. Then they sought to 
recover it in the limit when the cutoff is removed (see, for example, 
refs \cite{gauge}). In ref.\ \cite{TM1} one of us proposed a manifestly 
gauge invariant formulation of the ERG approach for pure Yang-Mills 
theories (see also refs \cite{TM2,TM3}). An important element 
of the formalism is an appropriate gauge invariant regularisation scheme.
As it was realised later in ref.\ \cite{TM3}, it amounts to the $SU(N|N)$ 
super-gauge extension of the original theory, with covariant higher 
derivatives, but spontaneously broken in the fermionic directions,
the resulting massive fields being Pauli-Villars regulating fields. 

The aim of the present contribution is to give a complete formulation 
of this extended theory and show that it is indeed free of ultraviolet 
divergences and is consistent in the continuum limit \ie when the 
cutoff $\Lambda \rightarrow \infty$. The article is organized as follows. 
In sec.\ 2 we explain the main idea of the regularisation scheme and 
describe in detail the structure of the $SU(N|N)$ super-gauge extension 
of the initial Yang-Mills theory. In sec.\ 3 we discuss the finiteness of the 
regularised theory. The full demonstration is quite long; in the present 
article we explain only basic points and present examples. Sec.\ 4 is devoted 
to the discussion of some potential problems arising due to the presence 
of a wrong sign massless vector field (loss of unitarity). We argue that 
in the continuum limit all the non-physical fields decouple and
unitarity in the physical sector is restored. A summary of results is 
presented in sec.\ 5.

\section{$SU(N|N)$ extension of $SU(N)$ Yang-Mills theory}

Consider the $D$-dimensional pure gauge theory with the gauge group $G$ and 
the action 
\beq
S_{YM} = \frac{1}{2} \int d^{D}x \, \tr \, \left( F_{\mu \nu} F^{\mu \nu}
\right). 
\label{S-YM0}
\eeq
Similar to the scalar case, as the first step of the construction 
of the regularised theory we modify the action by introducing 
the covariantised regulating function as follows: 
\beq
\hat{S}_{\Lambda} = \frac{1}{2} F_{\mu \nu} \{ c^{-1} \} F^{\mu \nu}.  
\label{S-YM1}
\eeq
For a given kernel $W$ and two functions $u(x)$ and $v(x)$ this
covariantisation (``wine'' \cite{TM1,TM2}) $u\{ W \} v$ is defined by 
\bea
& & u \{W\}v := \tr \int d^{D}x \, u(x) W \left(
-\frac{\nabla^{2}}{\Lambda^{2}} \right) \cdot v(x), \nonumber \\
& & \nabla_{\mu} = \partial_{\mu} - igA_{\mu}, \nonumber 
\eea 
where the dot means that $\nabla$ acts via commutation.

We would like to mention that the equation for the full effective action 
of the manifestly gauge invariant formulation of the ERG approach of ref.\ 
\cite{TM1} is  
\[
\Lambda \frac{\partial S_{\Lambda}}{\partial \Lambda} = 
- \frac{1}{2\Lambda^{2}} 
\frac{\delta S_{\Lambda}}{\delta A_{\mu}} \{ c' \}  
\frac{\delta \Sigma_{g}}{\delta A_{\mu}} + 
\frac{1}{2\Lambda^{2}} \frac{\delta}{\delta A_{\mu}} \{ c' \}  
\frac{\delta \Sigma_{g}}{\delta A_{\mu}},  
\]
where $\Sigma_{g} = g^{2}S_{\Lambda} - 2\hat{S}_{\Lambda}$. 

The $c( -\nabla^{2}/\Lambda^{2})$ regulator implements the higher 
covariant derivative regularisation proposed and developed in refs 
\cite{S1}, \cite{BS}. 
As it is known, however, some one-loop diagrams remain 
unregularised. A solution, proposed in ref.\ \cite{S2}, is to
supplement the scheme by Pauli-Villars regulating fields. 

{}From now on we will consider the case of the gauge group $G=SU(N)$.
Introducing bosonic and fermionic Pauli-Villars fields 
to cancel the one-loop ultraviolet divergences, one arrives at a certain 
system of regulating and physical fields. As discovered in 
ref.\ \cite{TM3}, for the purpose of regularisation it is enough and 
convenient to take the set of fields of the $SU(N|N)$ gauge theory. 
In order to generate the masses of the super-gauge fields without
destroying the gauge invariance we add 
a scalar Higgs field. This allows masses to be generated via the Higgs
mechanism, with the resulting heavy fields behaving precisely as
Pauli-Villars fields. As we will explain below, 
the unphysical fields decouple when $\Lambda \rightarrow \infty$, \ie 
when the regularisation is removed. 

To describe the $SU(N|N)$ extension of the theory (\ref{S-YM1}) with the
Higgs field let us consider first the graded Lie algebra of $SU(N|M)$ 
(see ref.\ \cite{Bar}). Its elements are given by Hermitian 
$(N+M) \times (N+M)$ matrices 
\[
{\cal H} = \left( \begin{array}{cc}
                   H_{1} & \theta \\
                   \theta^{\dagger} & H_{2}
                   \end{array}
            \right), 
\]
where $H_{N}$ ($H_{M}$) is an $N \times N$ ($M \times M$) Hermitian matrix with
complex bosonic elements, 
$\theta$ is an $N \times M$ matrix composed of complex Grassmann numbers, and
the matrices ${\cal H}$ are required to be supertraceless: 
\[
\str ({\cal H}) := \tr (H_{1}) - \tr (H_{2}) = 0.
\]
It is easy to see that the bosonic sector of the $SU(N|M)$ algebra forms the 
$SU(N) \times SU(M) \times U(1)$ subalgebra. 

We are interested in the case when $N=M$. In this case the $2N \times 2N$
identity  
matrix, $\one_{2N}$, is supertraceless and, therefore, is an element 
of the algebra. An arbitrary element ${\cal H}$ of $SU(N|N)$ can be written 
as 
\beq
 {\cal H} = {\cal H}^{0} \one_{2N} + {\cal H}^{A}T_{A},  \label{H-decomp}
\eeq
where $T_{A}$ are the other generators of $SU(N|N)$. They can be chosen both
traceless and supertraceless: $\str (T_{A}) = \tr (T_{A}) = 0$.
The index $A$ runs over $2(N^{2}-1)$ bosonic and $2N^{2}$ fermionic 
indices, and the Killing super-metric in the $T_A$ subspace,
\[
g_{AB} = \frac{1}{2} \str \left( T_{A} T_{B} \right), 
\]
is symmetric when both indices $A$ and $B$ are bosonic, antisymmetric 
when both are fermionic and is zero when one is bosonic and another is 
fermionic. 

Let us turn to the description of the $SU(N|N)$ gauge theory. 
The super-gauge field takes values in the graded Lie algebra 
of $SU(N|N)$ and, in accordance with eq.\ (\ref{H-decomp}), can be written as 
\[
{\cal A} = {\cal A}^{0}_{\mu} \one_{2N} + \tilde{\cal A}_\mu, 
\]
where
\[ 
\tilde{\cal A}_\mu = \left( \begin{array}{cc}
                   A^{1}_{\mu} & B_{\mu} \\
                   \bar{B}_{\mu} & A^{2}_{\mu}
                   \end{array}
            \right) = {\cal A}^{A}_{\mu} T_{A}. 
\]
The bosonic matrix $A^{1}_{\mu}$ is the physical field, \ie the 
field $A_{\mu}$ of the initial $SU(N)$ gauge theory. 
The fermionic field $B_{\mu}$ and the $SU(N)$ gauge field $A^{2}_{\mu}$
are part of the regulating structure. The pure Yang-Mills
part of the action of the extended theory is equal to 
\beq
{\cal }S_{YM} = \frac{1}{2} {\cal F}_{\mu \nu} \{ c^{-1}\} 
{\cal F}^{\mu \nu}, \label{S-YM2}
\eeq
where now the covariantisation is defined with the supertrace:
\beq
u \{W\}v := \str \int d^{D}x \, u(x) 
W \left( -\frac{\nabla^{2}}{\Lambda^{2}}\right) \cdot v(x) %%%\right]  
\eeq
and 
\bea
\nabla_{\mu} & = & \partial_{\mu} -ig{\cal A}_{\mu}, \nonumber \\
{\cal F}_{\mu \nu} & = & \partial_{\mu}{\cal A}_{\nu} - 
\partial_{\nu}{\cal A}_{\mu} -ig[{\cal A}_{\mu},{\cal A}_{\nu}], \nonumber 
\eea
and the regulating function $c^{-1}$ is chosen to be a polynomial in 
$(-\nabla^{2}/\Lambda^{2})$ of rank $n$. 

To generate masses for the fermionic Pauli-Villars fields via the Higgs 
mechanism we add a scalar sector. The super-scalar field is given by the 
super-matrix 
\[
{\cal C} = \left( \begin{array}{cc}
                   C^{1} & D \\
                   \bar{D} & C^{2}
                   \end{array}
            \right), 
\]
and we take the action of the scalar sector to be 
\beq
{\cal S}_{\cal C} = \nabla_{\mu} \cdot {\cal C} \{ \tilde{c}^{-1}\}
\nabla^{\mu} \cdot {\cal C} + 
\frac{\lambda}{4}\,\str\!  \int\! d^{D}x \, ({\cal C}^{2} - \Lambda^{2})^{2},
\label{S-C1}
\eeq
where the regulator $\tilde{c}^{-1}$ is a polynomial of rank $m$. 
The field ${\cal C}$ is not assumed to be supertraceless. 
Moreover, it acquires an expectation value which may be taken to be
$<{\cal C}> = \Lambda \sigma_{3}$, where 
\[
\sigma_{3} = \left( \begin{array}{cc}
                   \one_{N} & 0 \\
                   0 & -\one_{N}
                   \end{array}
            \right). 
\]
Expanding the super-scalar field around the stationary point of the 
potential in (\ref{S-C1}), \ie substituting 
${\cal C} \rightarrow \Lambda \sigma_{3} + {\cal C}$, one obtains the 
following expression for the action of the scalar sector: 
\bea
{\cal S}_{\cal C} & = & -g^{2} \Lambda^{2} [{\cal A}_{\mu},\sigma_{3}] \{
\tilde{c}^{-1}\} [{\cal A}_{\mu},\sigma_{3}] - 2ig \Lambda [{\cal
A}_{\mu},\sigma_{3}]  \{ \tilde{c}^{-1}\} \nabla^{\mu} \cdot {\cal C} \label{S-C2} \\
& + & \nabla_{\mu} \cdot{\cal C}\{ \tilde{c}^{-1}\} \nabla^{\mu} \cdot {\cal C} 
+ \frac{\lambda}{4} \; \str \int\! d^{D}x \, \left( \Lambda 
\{ \sigma_{3},{\cal C}\}_{+} + 
{\cal C}^{2} \right)^2.   \nonumber 
\eea
It is easy to see that the fields $B_{\mu}$, $C^{1}$ and 
$C^{2}$ acquire a mass of order $\Lambda$.  

Following 't Hooft's lead we choose the gauge fixing condition to be 
\[
\partial_{\mu} {\cal A}^{\mu} +ig \frac{\Lambda}{\chi}\tilde{c}^{-1} \hat{c}
[\sigma_{3},{\cal C}] = 0,  
\]
where $\hat{c}^{-1}(-\partial^{2}/\Lambda^{2})$ is a polynomial of rank $s$.
Note that it is not covariantised. The gauge fixing part of the action 
is equal to 
\bea
{\cal S}_{GF} & = & 
\chi \partial_{\mu} {\cal A}^{\mu} \cdot \hat{c}^{-1} \cdot 
\partial_{\nu} {\cal A}^{\nu} + 2ig \Lambda (\partial_{\mu} {\cal A}^{\mu}) 
\cdot \tilde{c}^{-1} \cdot [\sigma_{3},{\cal C}] \nonumber \\
& - & g^{2} \frac{\Lambda^{2}}{\chi} [\sigma_{3},{\cal C}] \cdot 
\tilde{c}^{-2} \hat{c} \cdot [\sigma_{3},{\cal C}], \label{S-GF}
\eea
where $u \cdot W \cdot v$ is defined as the supertrace analogue of eq.\ (\ref{fdWdg-def}). One 
can check that the last term in the action (\ref{S-GF}) gives a mass 
of order $\Lambda$ to the fermionic part of ${\cal C}$. 

The Faddeev-Popov ghost super-fields are defined as 
\[
\eta = \left( \begin{array}{cc}
                   \eta^{1} & \phi \\
                   \psi & \eta^{2}
                   \end{array}
            \right), 
\]
where the diagonal elements are fermionic while the off-diagonal ones 
are bosonic. The action of the ghost sector is given by  
\bea
{\cal S}_{ghost} & = & - \bar{\eta} \cdot \hat{c}^{-1}\tilde{c} 
\cdot \partial_{\mu} \nabla^{\mu} \cdot \eta  \label{S-gh} \\
& - & \int d^{D}x \; \str \left\{ \frac{\Lambda}{\chi} [\sigma_{3}, \bar{\eta}] 
\left( \Lambda [\sigma_{3},\eta] + [{\cal C},\eta] \right) \right\}.
\nonumber
\eea

\section{Finiteness of the regularised theory}

In the rest of the article we consider the case $D=4$. 

The complete set of Feynman rules will be given in ref.\ \cite{AKMT}.
To analyse the ultraviolet divergences in the theory we first calculate 
the superficial degree of divergence, ${\cal D}_{\Gamma}$, of 
a one-particle-irreducible diagram $\Gamma$ defined in the standard way. 
One can easily check that it is possible to choose the numbers $n$, $m$ 
and $s$, defining the ranks of the regulators, such that
all diagrams with two and more loops can be 
made finite. Among one-loop diagrams only those with $E_{\cal A} \leq 4$, 
where $E_{\cal A}$ denotes the number of external ${\cal A}$-lines, 
remain unregularised by this mechanism. However, they are finite 
due to the cancellation of the ultraviolet 
divergences between the contributions of the 
bosonic and fermionic propagators corresponding to internal lines. 
This cancellation will be referred to as the supertrace mechanism. 

To illustrate it, let us sketch the calculation of the one-loop 
diagram with two external ${\cal A}$-lines and two internal ${\cal A}$-lines. 
The terms of the perturbation theory expansion which generate this type of 
diagram, schematically omitting the Lorentz indices\footnote{Beware that
the commutators do not vanish once these are taken into account!}
involve the product of two vertices:
\beq
\str \left( [{\cal A}(x),{\cal A}(x)]{\cal A}(x) \right) %%%\wick  
\str \left( {\cal A}(y)[{\cal A}(y),{\cal A}(y)] \right),  \label{diag}
\eeq
where ${\cal A}$ stands for the gauge field or its derivative. 
The leading part of the 
propagator between the ${\cal A}^{A}$ and ${\cal A}^{B}$ fields 
in the momentum representation is proportional to $g^{AB}$. Using the 
completeness relation for the generators $T_{A}$ it is easy to show that 
by Wick pairing 
\[
%%\left<  %%%%an alternative!%%%%
\str ({\cal X}{\cal A}(x)) \str ({\cal A}(y){\cal Y})\wick{24}{14}
%%\right> 
= \left[ \frac{1}{2}\str ({\cal X}{\cal Y})  
- \frac{1}{4N} ( \tr  {\cal X} \;\str {\cal Y}
+ \str {\cal X} \; \tr  {\cal Y}) \right] \times \Delta (x-y),  
\]
where $\Delta (x-y)$ is a spacetime dependent 
factor coming from the propagator. Applying 
this formula to eq.\ (\ref{diag}) one can see that after the first pairing 
it reduces to
\beq
\frac{1}{2} 
\str \left([{\cal A}(x),{\cal A}(x)] [{\cal A}(y),{\cal A}(y)] \right) 
 \Delta (x-y). \label{diag1}
\eeq 
Here we have used the cyclicity property of the supertrace, $\str ({\cal
X}{\cal Y}) = \str ({\cal Y}{\cal X})$, which implies that $\str ([{\cal
X},{\cal Y}])=0$.  
For the next step we use the identity 
\beq
\str \left({\cal X} T_{A} {\cal Y} T^{A} \right) = 
\frac{1}{2}\str ({\cal X})\str({\cal Y}) 
- \frac{1}{4N} \left[\str ({\cal X} \sigma_{3}{\cal Y}) + 
str ({\cal X}{\cal Y} \sigma_{3}) \right], \label{splitid}
\eeq 
valid for any super-matrices ${\cal X}$ and ${\cal Y}$, which
follows from the already mentioned completeness relation. Using this 
identity we calculate the second ${\cal A}(x)$-${\cal A}(y)$ 
pairing in eq.\ (\ref{diag1}) and find
that the $\sigma_3$ terms generated by (\ref{splitid}) all cancel,
as they must -- to preserve the $SU(N|N)$ invariance, leaving only terms
of the form $\str{\cal A}\,\str{\cal A}$ or $\str{\cal A}\,\str\one$, 
both of which vanish because $\str {\cal A}=\str\one=0$.
This is a demonstration of the supertrace mechanism at work. 

One can check by direct calculation that the supertrace mechanism ensures 
the finiteness of all the diagrams with two and three external ${\cal
A}$-lines.  
For the diagrams with 4 external ${\cal A}$-lines the supertrace mechanism
is not sufficient (at finite $N$). However, these are already finite.
This follows because gauge invariant
effective vertices containing less than four ${\cal A}$s have already been
shown to be finite but gauge invariant effective vertices 
with a {\sl minimum} of four 
${\cal A}$s are already finite by power counting and
Ward identities for the $SU(N|N)$ gauge theory; this derivation is
given in ref.\ \cite{AKMT}.  

The rest of the one-loop diagrams and all the diagrams with more than one
loop are finite by the higher covariant derivative regularisation. 
To show this we first analyse the superficial degree of divergence 
%%${\cal D}_{\Gamma}$ 
of the one-loop diagrams with $E_{\cal A} > 4$ and 
other types of external lines, as well as the two-loop vacuum diagrams. 
It can be shown that they are finite if and only if the ranks of the 
regulating functions $c^{-1}$, $\tilde{c}^{-1}$ and $\hat{c}^{-1}$ satisfy 
the inequalities 
\bea
& & s > n > m > 0, \nonumber \\
& & n > 2, \; \; m > 1, \; \; n-m > 1. \label{nms-rel}
\eea
Then we show that these conditions are sufficient for the
superficial degree of divergence ${\cal D}_{\Gamma}$ of the rest of 
the one-particle-irreducible diagrams to be  
\[
{\cal D}_{\Gamma} < 0.
\]

This concludes the proof that all one-particle-irreducible 
diagrams in the $SU(N|N)$ theory with covariant 
higher derivatives whose action is given by eqs (\ref{S-YM2}), 
(\ref{S-C2}), (\ref{S-GF}) and (\ref{S-gh}) are ultraviolet 
finite.   

\section{Discussion of the unphysical sector}

The quadratic part of the action of the gauge sector of the $SU(N|N)$ 
theory is given by 
\beq
{\cal S}_{YM} = \int d^{D}x \, \frac{1}{2} \left[ \tr (F^{1}_{\mu \nu})^{2} - 
\tr (F^{2}_{\mu \nu})^{2} -  
2 \tr \left(\partial_{\mu} \bar{B}_{\nu} - \partial_{\nu} \bar{B}_{\mu})
(\partial_{\mu} B_{\nu} - \partial_{\nu} B_{\mu} \right) + \ldots \right]
   \label{S-YM3} 
\eeq
The appearance of the term with negative sign could potentially be a problem. 
At first glance it seems to be a signal of an instability. However, 
it is rather a sign of the loss of unitarity. 
This can already be seen in the example of $U(1|1)$ quantum mechanics
that we discuss now. 

Consider the Lagrangian of a simple harmonic potential: 
\[
L = \frac{1}{2} \str \dot{\cal X}^{2} - \frac{1}{2} \str {\cal X}^{2},
\]
where the Hermitian super-position ${\cal X}$ is given by the super-matrix 
\[
{\cal X} = \left( \begin{array}{cc}
                   x_{1} & \psi \\
                   \bar{\psi} & x_{2}
                   \end{array}
            \right).  
\]
The conjugate momenta are equal to 
\bea
& & p_{x_{1}} = \dot{x}_{1}, \; \; p_{x_{2}} = - \dot{x}_{2}, \; \;
[x_{j},p_{x_{j}}] = i, \nonumber \\
& & p_{\psi} = \dot{\bar{\psi}}, \; \; p_{\bar{\psi}} = - \dot{\psi}. \nonumber
\eea
Let us define the bosonic operators 
$a_{j} = (x_{j} + ip_{x_{j}})/\sqrt{2}$, 
$a_{j}^{\dagger} = (x_{j} - ip_{x_{j}})/\sqrt{2}$. They satisfy the commutation 
relations $[a_{i},a_{j}]=0$, $[a_{1}^{\dagger},a_{j}^{\dagger}]=0$, 
$[a_{i},a_{j}^{\dagger}]=\delta_{ij}$. 
The Hamiltonian of the system is equal to 
\[
H=\frac{1}{2} (a_{1}^{\dagger} a_{1} + a_{1}a_{1}^{\dagger}) - 
\frac{1}{2} (a_{2}^{\dagger} a_{2} + a_{2}a_{2}^{\dagger}) + \mbox{(fermionic
part)}.
\]
We proceed by introducing a complete set of states which includes the 
vacuum states $|0>_{1}$ and $|0>_{2}$ such that 
\[
a_{1}|0>_{1} = 0, \; \; \; a_{2}^{\dagger}|0>_{2} = 0, 
\]
and $n$-particle states 
\bea
& & |n>_{1} = \frac{1}{\sqrt{n!}} (a_{1}^{\dagger}
)^{n}|0>_{1},  \nonumber \\
& & |n>_{2} = \frac{1}{\sqrt{n!}}
(a_{2} )^{n}|0>_{2}.  \nonumber 
\eea
One can see that with these definitions 
$a_{2}$ plays the r\^{o}le of the creation operator of the particle of 
the second type. 
With these definitions the Hamiltonian is bounded from below. 
In particular 
\[
H|n>_{2} = + n|n>_{2}.
\]
Furthermore, it can be shown that it is these definitions that ensure that
the vacuum preserves the $U(1|1)$ symmetry.
However, the states $|n>_{2}$ with odd $n$ possess negative norms: 
\[
_{2}<n|n>_{2} = \frac{1}{n!} {_{2}<0|}(a_{2}^{\dagger})^{n}
(a_{2})^{n}|0>_{2} = (-1)^{n} {_{2}<0|0>_{2}}.
\]
This can be regarded as a violation of unitarity (negative probability). 
It may be mentioned that the appearance of negative norm states 
as a consequence of a wrong sign in part of the action is not that 
unusual. The Gupta-Bleuler quantization procedure 
relies on a modification of the Lagrangian which 
results in a wrong sign appearing in the $A^{0}$ part of the
action.\footnote{Of course, 
in the present case there is no analogue of the Gupta-Bleuler
condition.}

Transitions between the $A^{1}_{\mu}$-sector and $A^{2}_{\mu}$-sector 
are possible only via exchanges of fields with masses of order $\Lambda$. 
If $\Lambda$ is finite such transitions are possible, thus leading to a 
violation of unitarity in the $SU(N|N)$ gauge theory. 
In the continuum limit, \ie in the limit $\Lambda \rightarrow \infty$, 
all amplitudes for such transitions vanish. For example,  
the lowest order $A^{1} A^{2}$ amplitude appears from the term  
\beq
\str ({\cal A} {\cal A}) \, \str ({\cal A} {\cal A}) \times (\mbox{UV 
and IR finite
coefficient}). \label{AA-ampl}
\eeq
The requirement of gauge symmetry and dimensional considerations 
imply that (\ref{AA-ampl}) is in fact proportional to   
\[
\sim \int d^{4}x \, \frac{1}{\Lambda^{4}} \str ({\cal F}{\cal F}) \, \str ({\cal
F}{\cal F}) 
\]
and, therefore, vanishes as $\Lambda \rightarrow \infty$.

\section{Summary of the results}

In this article we have discussed   
$SU(N|N)$ gauge theory with higher covariant derivative
regulators, $c \left( -\nabla^{2}/\Lambda^{2}\right)$, and the Higgs field
viewed as a regularised extension of the $SU(N)$ pure Yang-Mills theory. 
Its structure is determined by the requirement that it can be 
used in the ERG equation presented in refs \cite{TM1} - \cite{TM3}. 

The extension includes the physical Yang-Mills field $A_{\mu}^{1} \equiv 
A_{\mu}$ of the initial theory and the regulating fields: 
the bosonic gauge field $A_{\mu}^{2}$, the fermionic Pauli-Villars field 
$B_{\mu}$ and the scalar Pauli-Villars fields $C^{i}$. 
We described the field content of the $SU(N|N)$ gauge theory 
and wrote down its action 
(see eqs (\ref{S-YM2}), (\ref{S-C2}), (\ref{S-GF}) and (\ref{S-gh})).  
All the regulator fields except $A_{\mu}^{2}$ acquire masses 
proportional to the momentum cutoff $\Lambda$ via the Higgs mechanism. 
The presence of the unphysical regulator fields lead to a source of
unitarity violation in the theory with finite cutoff.

We showed that in the four-dimensional case the 
one-loop one-particle-irreducible diagrams with two, three or four external 
${\cal A}$-lines are finite due to the supertrace mechanism. The rest of   
the one-loop one-particle-irreducible diagrams and all one-particle-irreducible 
diagrams with the number of loops $L \geq 2$ can be made finite by the proper 
choice of the regulating functions $c^{-1}$, $\tilde{c}^{-1}$ and 
$\hat{c}^{-1}$. The necessary and sufficient conditions on the parameters 
of these functions are given in (\ref{nms-rel}). 

When the regularisation is removed, \ie in the limit 
$\Lambda \rightarrow \infty$, the massive fields $B_{\mu}$ and
$C^{j}$ become infinitely heavy and decouple. As a consequence the physical 
sector, which is the original $SU(N)$ Yang-Mills theory, 
becomes decoupled from the unphysical sector of 
the field $A_{\mu}^{2}$. In this way the unitarity of the 
theory is restored in the continuum limit, \ie when 
$\Lambda \rightarrow \infty$. 

We expect the use of the $SU(N|N)$ regularised extension 
of the $SU(N)$ pure Yang-Mills theory to open up new possibilities 
of non-perturbative and gauge-invariant 
treatment of Yang-Mills theories in the framework of the ERG approach.
 
\section*{Acknowledgements}

{\tolerance=500 
We would like to thank M.Z. Iofa and J.I. Latorre for discussions and
valuable comments.  T.R.M. and Yu.K. acknowledge financial support from
the PPARC grant PPA/V/S/1998/00907, and Yu.K. acknowledges financial
support from the Russian Foundation for Basic Research (grant
00-02-17679). T.R.M. and S.A. acknowledge support from PPARC SPG
PPA/G/S/1998/00527. J.F.T. thanks PPARC for support through a
studentship.

\end{document}